\RequirePackage{fix-cm}
\documentclass[onecollarge]{svjour3}   
\smartqed  % flush right qed marks, e.g. at end of proof
\usepackage{graphicx}
\usepackage{slashed}
\begin{document}

\title{Quark mass functions and pion structure in the Covariant Spectator Theory%\thanks{Grants or other notes
%about the article that should go on the front page should be
%placed here. General acknowledgments should be placed at the end of the article.}
}
%\subtitle{Do you have a subtitle?\\ If so, write it here}

%\titlerunning{Short form of title}        % if too long for running head

\author{Elmar P. Biernat         \and
       Franz Gross \and 
       Teresa Pe\~na \and 
       Alfred Stadler \and
       Sofia Leit\~ao
       %etc.
}

%\authorrunning{Short form of author list} % if too long for running head

\institute{Elmar P. Biernat, Teresa Pe\~na, Alfred Stadler, Sofia Leit\~ao \at 
Centro de F\' isica Te\' orica de Part\' iculas, Instituto Superior T\'ecnico, Av.\ Rovisco Pais, 
1049-001 Lisboa, Portugal\\ 
              \email{elmar.biernat@tecnico.ulisboa.pt}           %  \\
%             \emph{Present address:} of F. Author  %  if needed
          \and Alfred Stadler \at 
          Departamento de F\'isica, Universidade de \'Evora, 7000-671 \'Evora, Portugal
          \and
          Franz Gross \at Thomas Jefferson National Accelerator Facility (JLab), Newport News, VA 23606, USA and College of William and Mary, Williamsburg, Virginia 23188,
USA}

\date{Received: date / Accepted: date}
% The correct dates will be entered by the editor

\maketitle

\begin{abstract}
The Covariant Spectator Theory is applied to the description of quarks and the pion. The dressed quark mass function is calculated dynamically in Minkowski space and used in the calculation of the pion electromagnetic form factor. The effects of the mass function on the pion form factor and the different quark-pole contributions to the triangle diagram are analyzed.
% \subclass{MSC code1 \and MSC code2 \and more}
\end{abstract}

\section{Introduction}
\begin{figure}
% Use the relevant command to insert your figure file.
% For example, with the graphicx package use
   \includegraphics[clip=10cm,width=0.5\textwidth]{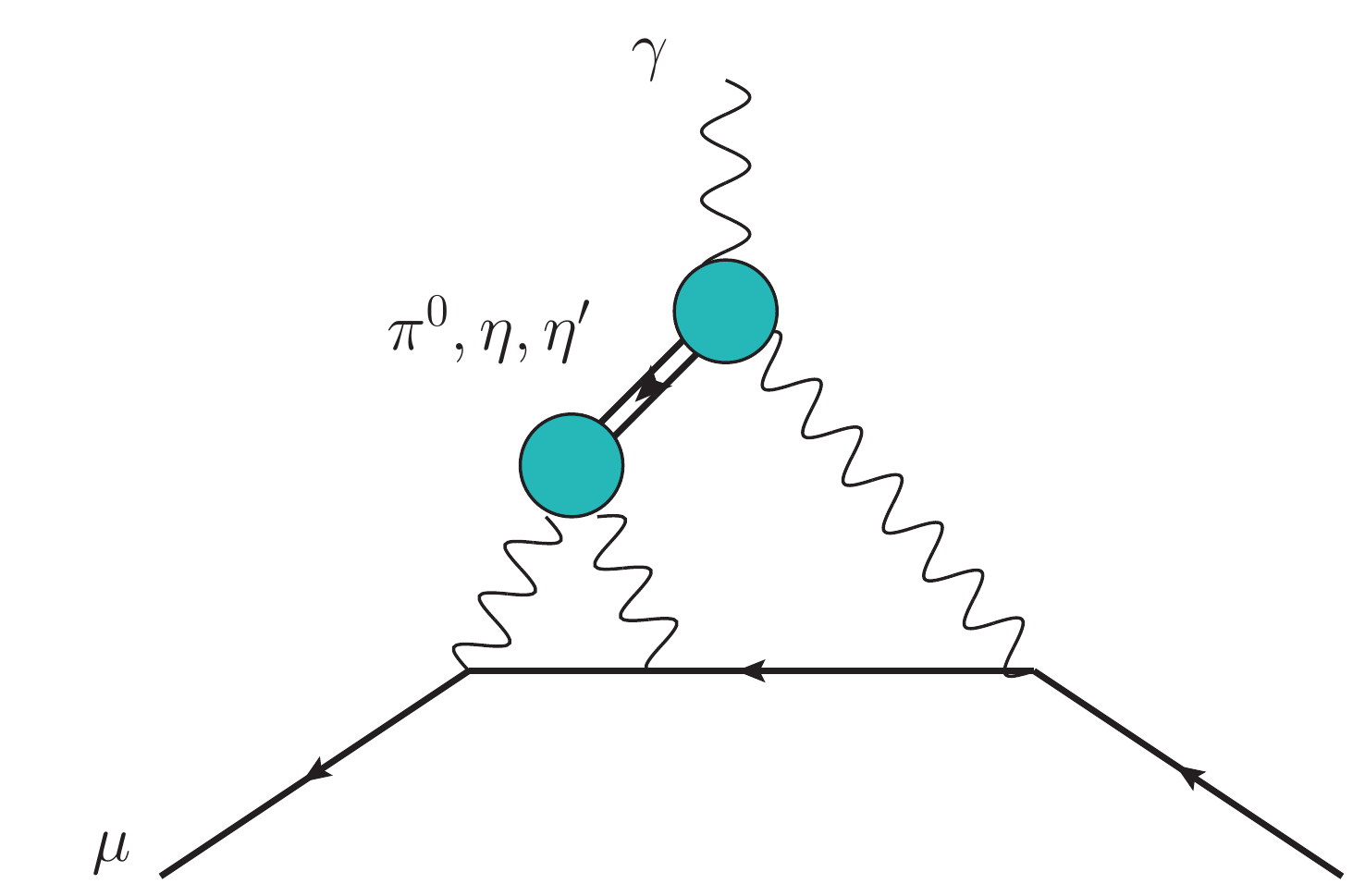} \includegraphics[clip=10cm,width=0.5\textwidth]{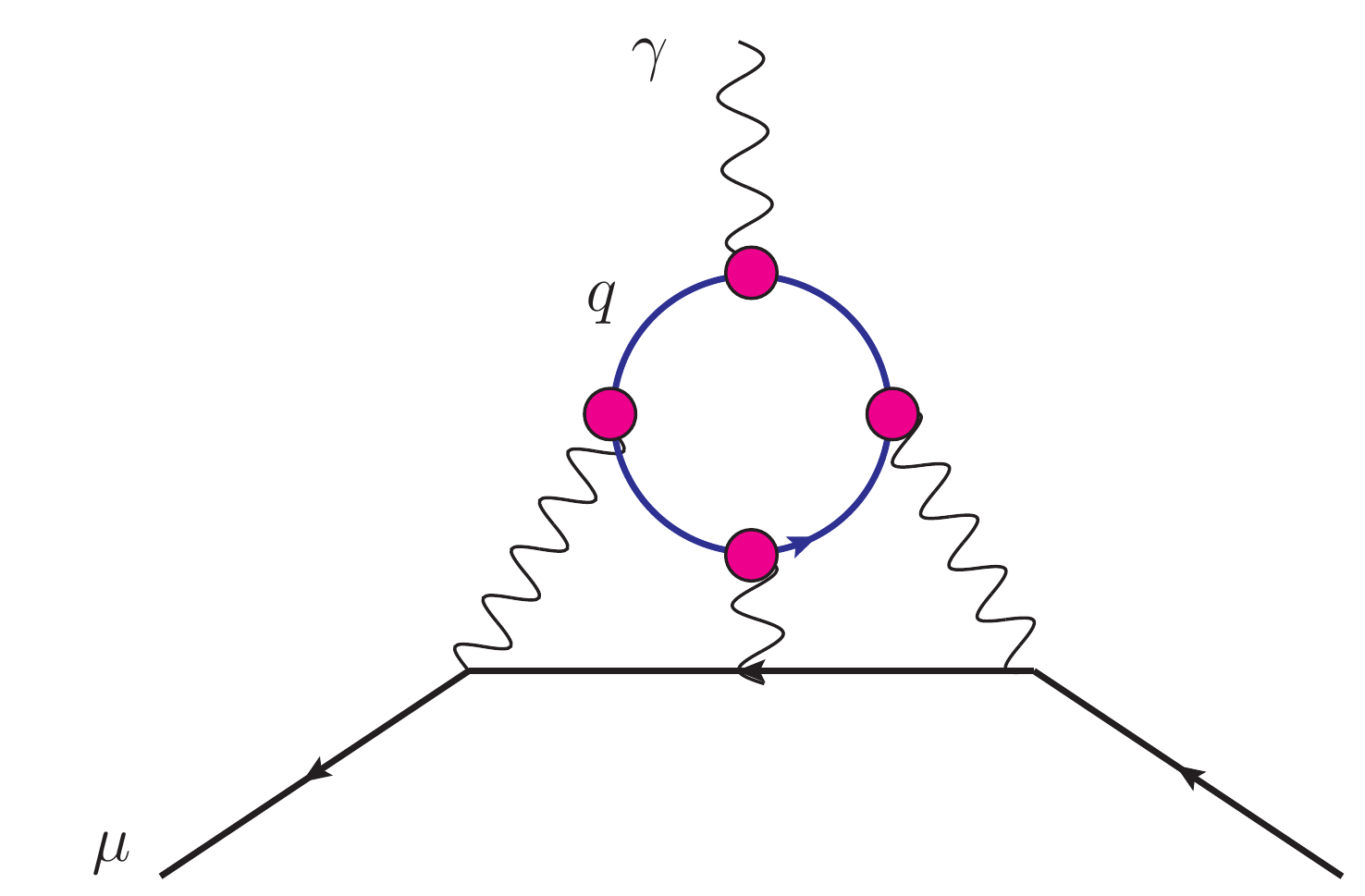}
  
% figure caption is below the figure
\caption{The pseudoscalar-meson-pole contribution (left panel) with the meson transition form factors (turquoise blobs) and the dressed-quark-loop contribution (right panel) with the dressed quark-photon vertices (pink blobs) and the dressed quark propagators (blue lines).}
\label{fig:1}       % Give a unique label
\end{figure}

Meson form factors are important physical observables that contain the information on the structure of a meson, and various physical processes and quantities depend on them. A prominent example that has recently attracted much attention is the muon anomalous magnetic moment, a low-energy observable both measured and calculated to extremely high precision. Its present experimental value persistently deviates from the theoretical prediction by more than 3$\sigma$, a discrepancy often interpreted as a signature of \lq \lq physics beyond the Standard Model''. Therefore, in order to draw definite conclusions, it is of highest importance to increase the accuracy of both measurement and theoretical calculations. The largest theoretical uncertainty come from the hadronic vacuum polarization and hadronic light-by-light scattering contributions. For the latter, the most relevant contributions are the pseudoscalar-meson-pole diagrams related to the meson transition form factors, and the dressed-quark-loop diagram, all of which depend on the dressed quark propagator and the dressed quark-photon vertex. These contributions to the muon anomaly are diagrammatically depicted in Fig.~\ref{fig:1}

This article focuses on the dressed quark propagators (i.e. the dressed quark mass function), the dressed quark-photon vertex, and the meson vertex functions as the ingredients for the calculation of meson electromagnetic form factors. In particular, we concentrate here on the pion electromagnetic form factor in the spacelike region of momentum-transfer squared $Q^2>0$. The pion is the ideal test case for relativistic calculations because of its simplicity and the large amount of experimental data available from various hadron facilities worldwide, such as the Jefferson Lab and FAIR-GSI.

Various theoretical methods have been developed in order to describe the non-perturbative dynamics underlying the pion and other hadronic bound states. For instance, lattice QCD~\cite{LatticeQCD}, Hamiltonian approaches based on Dirac's forms of dynamics~\cite{LFQFT}, as well as methods based on the Dyson-Schwinger/Bethe-Salpeter (DSBS) approach and the mass gap equation~\cite{DSBS} have made significant contributions to our understanding of hadrons.

Our theoretical framework is the Covariant Spectator Theory (CST)~\cite{Gro69}, a nonperturbative approach with the unique feature that it allows the use of confining interaction kernels with a Lorentz scalar structure without violating chiral symmetry. Similar to the DSBS approaches, it incorporates both relativistic covariance and dynamical chiral-symmetry breaking, features that are indispensable for the description of the light mesons, such as the pion, $\eta$, and $\eta'$. In contrast to lattice QCD and the DSBS approaches, the CST equations are solved in Minkowski space, which allows, for instance, for a straightforward extension of form factor results from the spacelike- to the timelike-$Q^2$ region. This region is relevant for medium effects searched in experiments to study matter under extreme conditions.

\section{Quark mass functions in CST}
Dynamical chiral-symmetry breaking implies the dynamical generation of a constituent quark mass via the strong interaction. In the charge-conjugation-invariant formulation of CST~\cite{Savkli:1999me}, this is realized by the CST Dyson equation given by
\begin{equation}
 S^{-1}(p)=S_0^{-1}( p)+\frac12 Z_R
\sum_{\sigma=\pm} \int \frac{\mathrm d^3 k}{(2\pi)^3}\frac{m}{E_k} \mathcal V(p, \hat k_\sigma)\Lambda (\hat{\slashed k}_\sigma)\,,\label{eq:CSTDE}
\end{equation}
where $S( p)=Z(p^2)\left[M(p^2)-\slashed p -\mathrm i\epsilon \right]^{-1}$ and $S_0(p)=\left[m_0-\slashed p -\mathrm i\epsilon \right]^{-1}$ are the dressed and bare quark propagators, respectively, with $p$ the off-shell quark four-momentum, $M(p^2)=Z(p^2)[A(p^2)+m_0]$ the dressed quark mass function, $A(p^2)$ the scalar part of the quark self-energy, $Z(p^2)$ the wave-function renormalization, $m_0$ the bare quark mass, $m$ the (dressed) constituent quark mass, $Z_R\equiv Z(m^2)/(1-2m M'(m^2))$ a constant with $M'\equiv \mathrm dM/\mathrm dp^2$, $\mathcal V (p,\hat k_\pm)$ the strong-interaction kernel, $\Lambda({k}) =(M(k^2)+ \slashed {k })/2M(k^2)$, and $\hat {k}_\pm=(\pm E_k,\vec{k})$ is the on-shell quark four-momentum with $E_k=\sqrt{m^2+\vec k^2}$. The constituent quark mass $m$ is determined as the value of the mass function where $S$ has a real pole, i.e.  $m\equiv M(m^2)$. Equation~(\ref{eq:CSTDE}) describes the dynamical generation of the dressed quark mass function due to its dressing through the interaction kernel. In the present work we use a kernel of the form
\begin{equation}
 \mathcal V(p,\hat k_\pm)=(\mathbf{1}\otimes\mathbf{1}+\gamma^5\otimes\gamma^5) V_L(p,\hat k_\pm )+\gamma^{\mu} \otimes\gamma_\mu h^2(p^2) \frac{C}{2m} (2\pi)^3E_k \delta^3(\vec p-\vec k)\,,  \label{eq:V}
\end{equation}
where $V_L(p,\hat k_\pm)$ is a covariant generalization in momentum space of the linear confining potential, satisfying the condition~\cite{Gro69}
\begin{eqnarray}
 \int \frac{\mathrm d^3 k}{E_k} V_{L} (p,\hat k)=0\,,
 \label{eq:VLzero}
\end{eqnarray}
which is the CST version of the property that describes the vanishing of the nonrelativistic linear potential at the origin in coordinate space.
The last term of the kernel of Eq.~(\ref{eq:V}) is a covariant generalization in momentum space of a constant potential $C$ in coordinate space, and $h(p^2)$ is a strong quark form factor. In Ref.~\cite{PhysRevD.90.096008} we proved that the kernel of Eq.~(\ref{eq:V}) satisfies the axial-vector Ward-Takahashi identity which ensures dynamical chiral-symmetry breaking. This works with a mixed (equal-weight) Lorentz pseudoscalar-scalar linear confining kernel since such a kernel does not contribute to the CST Dyson equation (\ref{eq:CSTDE}) for the dressed quark propagator because of the property of Eq.~(\ref{eq:VLzero}). After solving Eq.~(\ref{eq:CSTDE}) analytically in the chiral limit ($m_0=0$), where $Z_RC\rightarrow m$,  the dynamical quark mass function is given by
\begin{eqnarray}
 M(p^2)=m h^2(p^2)\,. 
 \label{eq:massfunction}
\end{eqnarray}

\begin{figure}[htb]
\centerline{%
\includegraphics[width=0.75\textwidth]{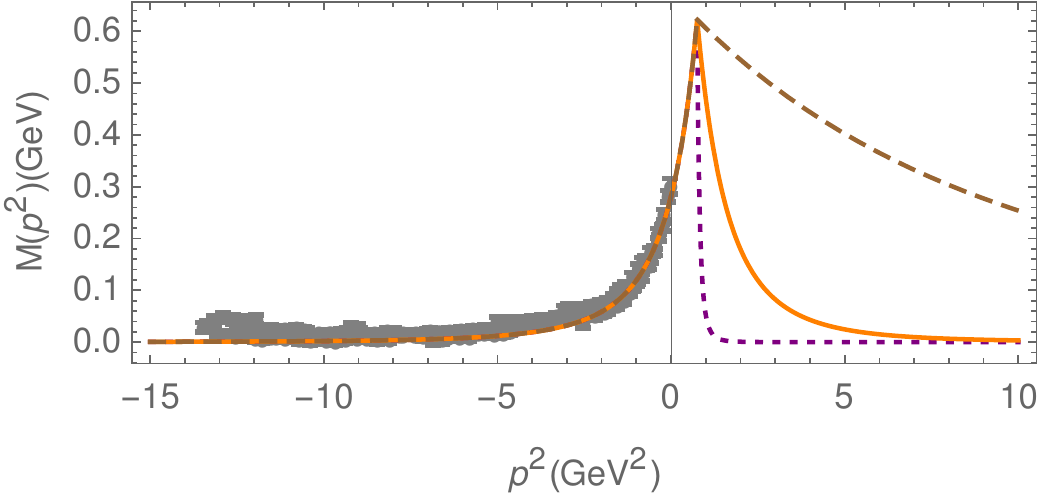}}
\caption{The chiral-limit mass function compared with the lattice QCD data~\cite{Bowman:2005vx} (extrapolated to the chiral limit). In the timelike region 3 different possible shapes are shown.}
\label{Fig:massfunction}
\end{figure}

The strong quark form factor $h(p^2)$ depends on the chiral-limit constituent quark mass $m$ and a mass cutoff parameter $\Lambda$. Both these parameters are determined by a fit of the mass function Eq.~(\ref{eq:massfunction}) at negative momenta-squared $p^2$ to the the available lattice-QCD data~\cite{Bowman:2005vx} (extrapolated to the chiral limit). Since for timelike momenta $p^2>0$ there are no lattice QCD data available, we adopt an Ansatz for $h(p^2)$ in this region. Varying the shape of $h$ for $p^2>0$ will allow us to study the sensitivity of the pion form factor calculation on the quark mass function. Figure~\ref{Fig:massfunction} compares the mass function with the lattice QCD data in the chiral limit and also shows possible shapes in the timelike region.
\section{The pion electromagnetic form factor in CST}

\begin{figure}[htb]
\centerline{%
\includegraphics[width=0.75\textwidth]{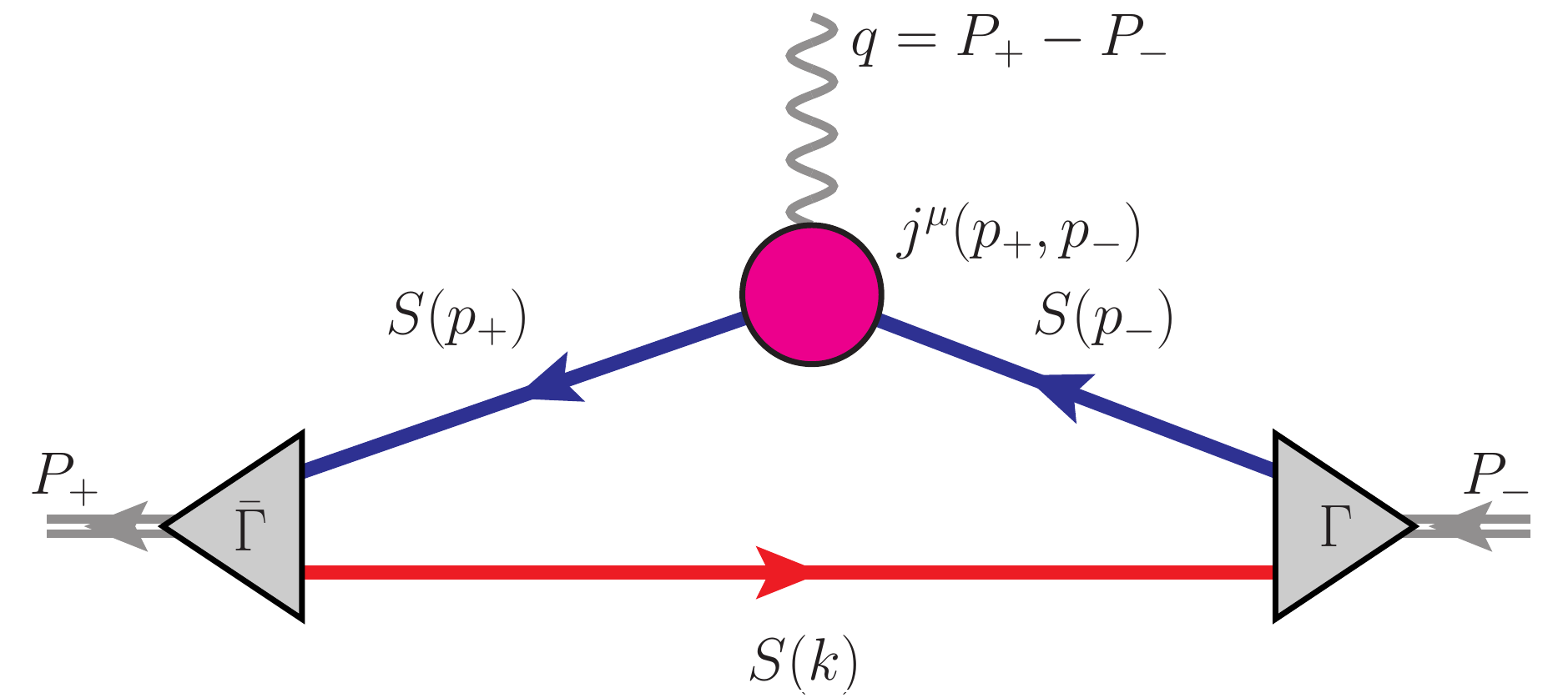}}
\caption{The triangle diagram for the pion current which describes the coupling of the virtual photon (wiggly line) to the the quark (blue line), with the antiquark as a spectator (red line).}
\label{Fig:TriangleA}
\end{figure}

Next we use the quark mass function in the calculation of the pion electromagnetic form factor. In impulse approximation, the electromagnetic pion current is calculated from the sum of two triangle diagrams in which the virtual photon couples either to the quark or to the antiquark. One of these diagrams is shown in Fig.~\ref{Fig:TriangleA}. 
The pion current is given by
\begin{eqnarray}
F(Q^2) (P_++P_-)^\mu&=& \mathrm i \int \frac{\mathrm d^4k}{(2\pi)^4} \mathrm {Tr} \left[\bar \Gamma (k, p_+) S(p_+)j^\mu (p_+,p_-) S(p_-)\right.\nonumber\\&&\times\left.\Gamma (p_-,k) S(k)\right] \,,\label{eq:pioncurrent}
\end{eqnarray}
where $F(Q^2)$ is the pion electromagnetic form factor and\linebreak ${\sloppy Q^2=-q^2=-(P_+-P_-)^2}$ is the virtuality of the photon.
According to the charge-conjugation-invariant CST prescription of how to perform the $k_0$ contour integration in Eq.~(\ref{eq:pioncurrent}) one takes all six quark propagator-pole contributions into account. These are the two positive- and negative-energy spectator quark poles at $k^2=m^2$ and the four positive- and negative-energy poles at $p_\pm^2=m^2$ of the active quark to which the photon couples~\cite{Biernat:2015xya}. 

For the dressed off-shell quark-photon vertex $j^\mu$ we use the current given by~\cite{Biernat:2015xya} 
\begin{eqnarray}
 j^\mu(p_+,p_-)&=& h(p_+^2)\left[f(p_+,p_-)\gamma^\mu +\delta(p_+,p_-) \Lambda (-p_+)\gamma^\mu\right.\nonumber\\&&+\left.\delta(p_-,p_+) \gamma^\mu\Lambda (-p_-)+g(p_+,p_-) \Lambda (-p_+)\gamma^\mu\Lambda (-p_-)\right]h(p_-^2)\,,\nonumber\\
\end{eqnarray}
where $f$, $\delta$, and $g$ are scalar functions determined through the Ward-Takahashi identity in terms of the $h$ form factors, according to the Gross-Riska prescription~\cite{Gro87} to ensure gauge invariance. For the pion vertex function $\Gamma$, we take a simple Ansatz near the chiral limit of the form $\Gamma (p_-,k)\propto h(p_-^2)h(k^2)\gamma^5$ and $\bar \Gamma (k,p_+)\propto h(p_+^2)h(k^2)\gamma^5$~\cite{Biernat:2015xya}.
\section {Results and conclusions}
\begin{figure}[htb]
\centerline{%
\includegraphics[width=0.75\textwidth]{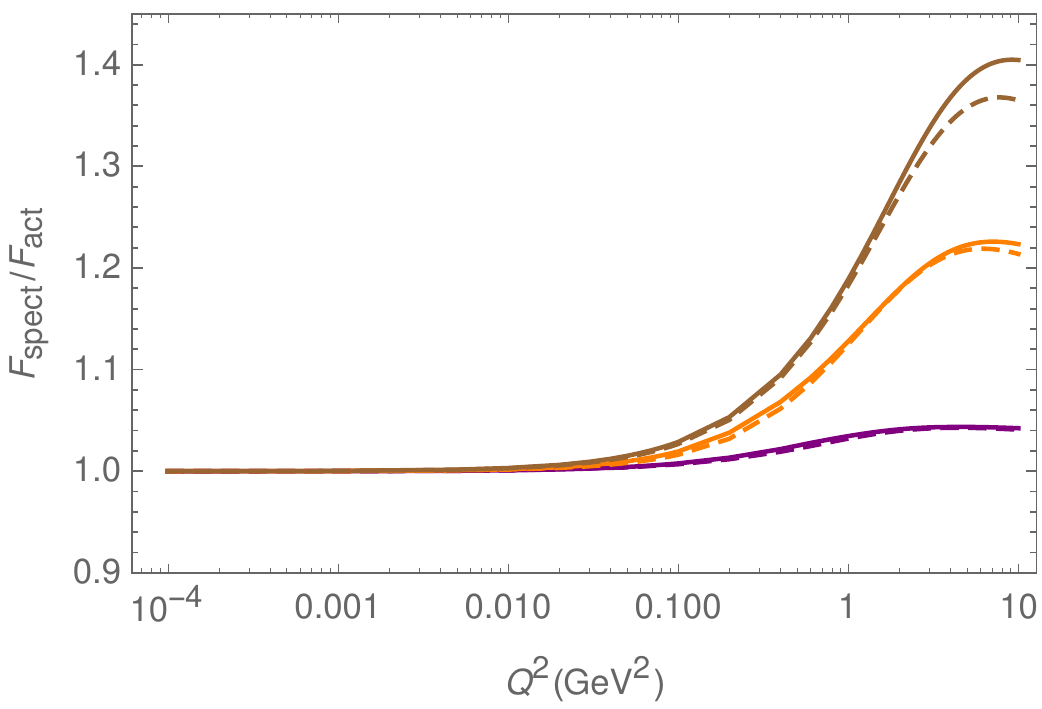}}
\caption{The ratio $F_{\rm spect}/F_{\rm act}$ calculated with constant quark masses (dashed lines)
and momentum-dependent quark mass functions (solid lines), and different values of the pion mass $\mu$.
The pairs of curves, from top to bottom, are the results obtained
with $\mu=0.6$ (brown), $0.42$ (orange), and $0.14$ GeV (purple).}
\label{Fig:SpectOverAct_running_vs_fixed}
\end{figure}

\begin{figure}[htb]
\centerline{%
\includegraphics[width=0.75\textwidth]{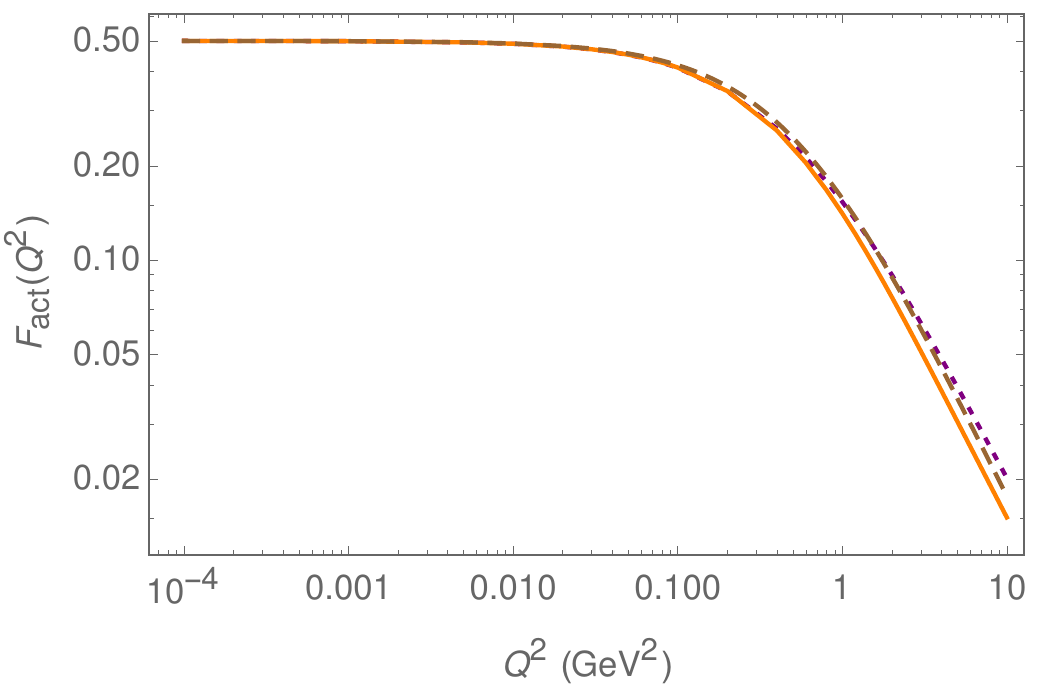}}
\caption{$F_{\rm act}$ when calculated with different quark mass functions, corresponding to the ones shown in Fig.~\ref{Fig:massfunction}.}
\label{Active_Poles_dynamical_mu_420_var_alpha}
\end{figure}

In Fig.~\ref{Fig:SpectOverAct_running_vs_fixed} the ratio $F_{\rm spect}/F_{\rm act}$  versus $Q^2$ is shown, where $F_{\rm spect}$ and $F_{\rm act}$ are the spectator and active quark-pole contributions to the pion electromagnetic form factor, respectively. Also shown are the results obtained with constant quark masses and quark mass functions in order to analyse the effect of the momentum dependence of the quark mass in the pion form factor. 
In Fig.~\ref{Active_Poles_dynamical_mu_420_var_alpha} the results of $F_{\rm act}$ when calculated with different mass functions in the timelike region (see Fig.~\ref{Fig:massfunction}) are compared. 
Note that the computation of $F_{\rm spect}$ tests the mass function only in the spacelike region where it is fixed by the lattice QCD data, hence all curves coincide in this case.

One conclusion of this work can be drawn from Fig.~\ref{Fig:SpectOverAct_running_vs_fixed}: For small pion masses $\mu$, the active quark contributions $F_{\rm act}$ are as important as the spectator contributions $F_{\rm spect}$, over the whole range of $Q^2$, and only for large $\mu$ and large $Q^2$, $F_{\rm act}$ is  suppressed as compared to $F_{\rm spect}$ by about 30\%. This suppression is slightly stronger for momentum-dependent than for constant quark masses. For small $\mu$, $F_{\rm act}$ and $F_{\rm spect}$ are nearly identical, not only in magnitude but also in shape, even for large $Q^2$. Furthermore, we find that the pion form factor in this model is surprisingly insensitive to the functional form of the strong quark form factors and quark mass function, as can be seen in Fig.~\ref{Active_Poles_dynamical_mu_420_var_alpha}.

Despite its simplicity, this first calculation of the pion form factor in the charge-conjugation-invariant CST showed the importance of the different quark-pole contributions to the triangle diagram and its insensitivity to the choice of the time-like dependence of the strong quark form factors. It will be the reference work for future form factor calculations within this framework.

\begin{acknowledgements}
This work was supported by Funda\c c\~ao para a Ci\^encia e a 
Tecnologia (FCT) under Grants No. CFTP-FCT (UID/FIS/00777/2013), No. CERN/FP/123580/2011, No. SFRH/BPD/100578/2014, and No. SFRH/BD/92637/2013, and by the European Community's Seventh Framework Programme FP7/2007-2013 under Grant Agreement No.\ 283286. F.G. was supported by the U.S. Department of Energy, Office of Science, Office of Nuclear Physics under contract DE-AC05-06OR23177.
\end{acknowledgements}

\end{document}